\documentclass{PoS}

\usepackage{graphicx} 
\usepackage{amssymb} 
\usepackage{amsmath} 
\usepackage{lineno}
\usepackage{tabularx}
\usepackage[babel=true]{csquotes}    
\usepackage{wrapfig}

\def\beq     {\begin{equation}}
\def\eeq     {\end{equation}}

\newcommand{\cf}[1]{{Fig.~\ref{#1}}}
\newcommand{\ct}[1]{{Table~\ref{#1}}}

\newcommand{\eg}{{\it e.g.}}
\newcommand{\etal}{{\it et al.}}

\def\lsim{\raise0.3ex\hbox{$<$\kern-0.75em\raise-1.1ex\hbox{$\sim$}}}
\def\gsim{\raise0.3ex\hbox{$>$\kern-0.75em\raise-1.1ex\hbox{$\sim$}}}

\def\pA   {$pA$}

\def\PbA {Pb$A$}
\def\jpsi    {\mbox{$J/\psi$}}
\def\upsi    {\mbox{$\Upsilon$}}

\title{Studying the high {\boldmath $x$} frontier with A Fixed-Target ExpeRiment at the LHC}

\ShortTitle{Studying the high $x$ frontier with AFTER@LHC}

\author{\speaker{A.~Rakotozafindrabe}\\
        IRFU/SPhN, CEA Saclay, 91191 Gif-sur-Yvette Cedex, France
        }        

\author{M.~Anselmino, R.~Arnaldi, E.~Scomparin\\
        Dip. di Fisica and INFN Sez. Torino, Via P. Giuria 1, I-10125, Torino, Italy
        }
\author{S.J.~Brodsky\\
        SLAC National\,Accelerator\,Laboratory, Stanford University, Menlo Park, CA 94025,USA
        }
\author{V.~Chambert, J.P. Didelez, B. Genolini, C.~Hadjidakis, J.P.~Lansberg, C.~Lorc\'e, P.~Rosier \\
        IPNO, Universit\'e Paris-Sud, CNRS/IN2P3, F-91406, Orsay, France
        }
\author{E.G.~Ferreiro\\
        Dept. de F{\'\i}sica de Part{\'\i}culas, USC, 15782 Santiago de Compostella, Spain
        }
\author{F.~Fleuret\\
        LLR, \'Ecole Polytechnique, CNRS/IN2P3,  91128 Palaiseau, France
        }
\author{I.~Schienbein\\
        LPSC, Univ. Joseph Fourier, CNRS/IN2P3/INPG, 38026 Grenoble, France
        }
\author{U.I.~Uggerh\o j\\
        Department of Physics and Astronomy, University of Aarhus, Denmark
        }        

\abstract{The opportunities which are offered by a next generation 
and multi-purpose fixed-target experiment exploiting the proton and lead LHC 
beams extracted by a bent crystal are outlined. In particular, such an experiment can 
greatly complement facilities with lepton beams by unraveling 
the partonic structure of polarised and unpolarised 
nucleons and of nuclei, especially at large momentum fractions.
}

\FullConference{XXI International Workshop on Deep-Inelastic Scattering and Related Subject -DIS2013,\\
		22-26 April 2013\\
		Marseille, France}

\begin{document}

\section{Introduction}
\label{sec:intro}

The Large Hadron Collider (LHC) has been providing the most energetic beams ever, 
delivering proton beams at $3.5$~TeV and 4~TeV, and lead beams at $1.38$~TeV per nucleon. 
Projects of detector and accelerator upgrades are being presently discussed, and the idea of 
colliding the high energy LHC beams with an electron beam is being investigated by the LHeC study 
group~\cite{AbelleiraFernandez:2012cc}. 

The LHC complex could be further exploited in a very cost-effective way by recycling 
the beam loss towards A multipurpose Fixed-Target ExpeRiment, named AFTER~\cite{Brodsky:2012vg}. 
A bent crystal can be used to extract a fraction of the beam loss~\cite{Uggerhoj:2005xz}. 
Intensity as high as  $5 \times 10^8 \, p^+ s^{-1}$ can certainly be obtained. 
This technology was successfully tested at SPS~\cite{Arduini:1997kh}, Fermilab~\cite{Asseev:1997yi}, Protvino~\cite{Afonin:2012zz}
for proton and recently for lead beams at SPS~\cite{Scandale:2011za}. 
It has been proposed as a smart  alternative for the upgrade of the LHC collimation 
system and will be tested by the LUA9 collaboration in the years to come~\cite{LUA9-letter-of-intent}.
The available energy in the c.m.s.\ amounts to $\sqrt{s_{NN}} = 72$~GeV in \PbA\ collisions and 
$\sqrt{s_{NN}} = 115$~GeV in \pA\ collisions. AFTER will benefit from the typical advantages of 
a fixed-target experiment, notably the outstanding luminosities (\ct{tab:lumi}), the high boost between 
the laboratory and the center-of-momentum frame ($\gamma \simeq 60$ with the 7 TeV proton beam), the versatility of the target species, 
and the possibility to polarise the target. 

\begin{table}[hbt]
\begin{tabular}{ccccccc}
\hline
Beam
 & Target
 & Thickness (\rm{cm})
 & $\rho$ (\rm{g cm}$^{-3}$)
 & $A$
 & $\mathcal L$ ($\mu$\rm{b}$^{-1}$ \rm{s}$^{-1}$)  
 & $\int\mathcal L$ (\rm{pb}$^{-1}$ \rm{y}$^{-1}$)  \\
\hline
$p$ & Solid H &10 & 0.088 & 1  & 260 & 2600\\
$p$ & Liquid H &100 & 0.068 & 1  & 2000 & 20000\\
$p$ & Liquid D & 100 & 0.16 & 2  & 2400 & 24000\\
$p$ & Pb & 1&11.35 & 207  & 16 & 160\\
\hline
Pb & Solid H &10& 0.088 & 1  & 0.11 & 0.11\\
Pb & Liquid H &100 & 0.068 & 1  & 0.8 & 0.8\\
Pb & Liquid D &100& 0.16 & 2  & 1 & 1\\
Pb & Pb & 1&11.35 & 207  & 0.007 & 0.007\\
\hline
\end{tabular}
\caption{Integrated luminosities per year ($10^7$s for $p$ and $10^6$s for Pb) obtained 
with an extracted beam of $5\times 10^8$ $p^+$/s ($3.5$ TeV) and of $2\times 10^5$ Pb/s ($1.38$ TeV) 
for various targets. The expected yearly luminosities should be way above that of 
RHIC in the same energy range.}
\label{tab:lumi}
\end{table}

Thanks to the very energetic LHC beams, AFTER will grant access to the target rapidity region i.e.\ 
the full backward region, up to $x_F \rightarrow -1$ or equivalently $x_2 \rightarrow 1$, 
which is largely uncharted. The $x$ window covered by AFTER will be complementary~\cite{Brodsky:2012vg} to those 
covered by COMPASS, RHIC and their upgrades, which do not go beyond $x \sim 0.3$, and to 
electron-ion facilities (such as LHeC) that will access the very low values of $x$, below $10^{-2}$. 
Recently, the importance of large $x$ physics has been emphasized in~\cite{Accardi:2013pra}. 
As an example, the \cf{fig:eps09} shows the current limitations of our knowledge on the nuclear 
PDF (nPDF). It is worth underlining that the uncertainties at high $x$ are larger than at low $x$ and 
that they do not decrease with increasing scale, contrary to what is observed at low $x$. In the following, 
we will discuss a short selection\footnote{More details can be found in~\cite{Brodsky:2012vg,Lansberg:2012kf}.} 
of opportunities offered by AFTER for the physics at high $x$.

\begin{figure}
 \begin{minipage}[b]{.55\linewidth}
  \centering\includegraphics[width=\textwidth]{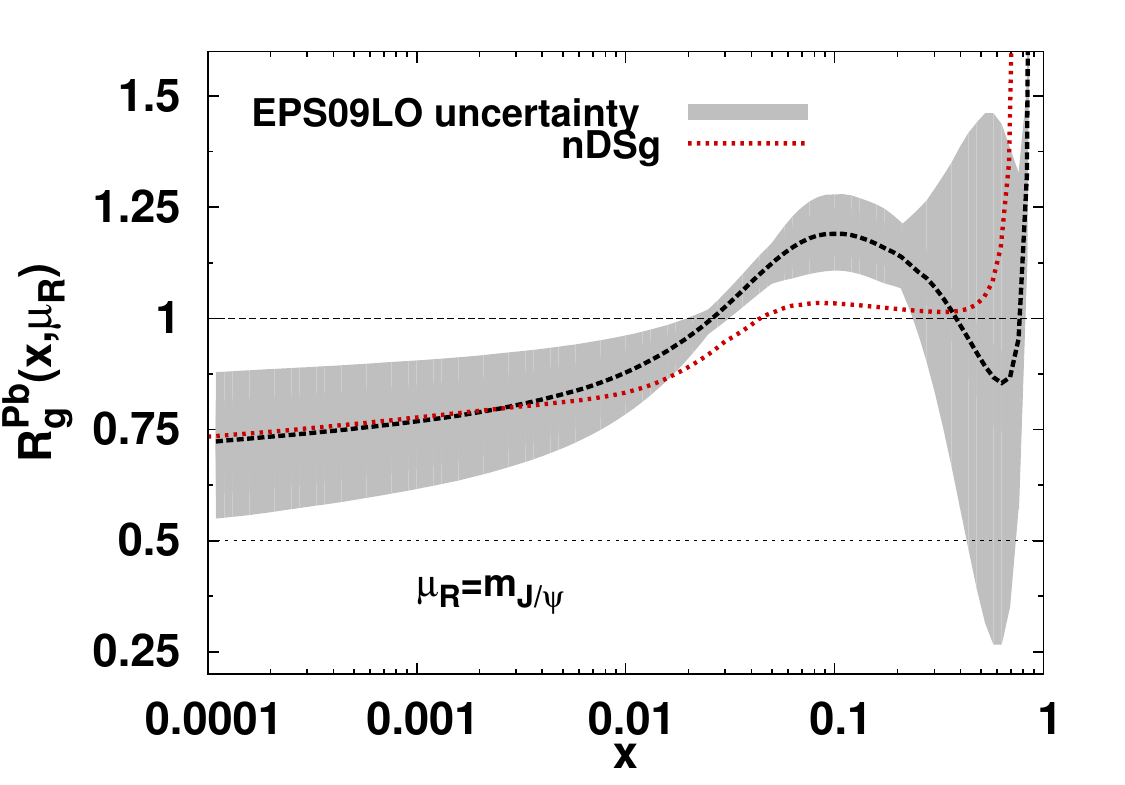} 
  \caption{The uncertainty on the nuclear gluon PDF in Pb at the scale of the \jpsi\ mass. \label{fig:eps09}}
 \end{minipage} \hfill
 \begin{minipage}[b]{.4\linewidth}
  \centering\includegraphics[width=\textwidth]{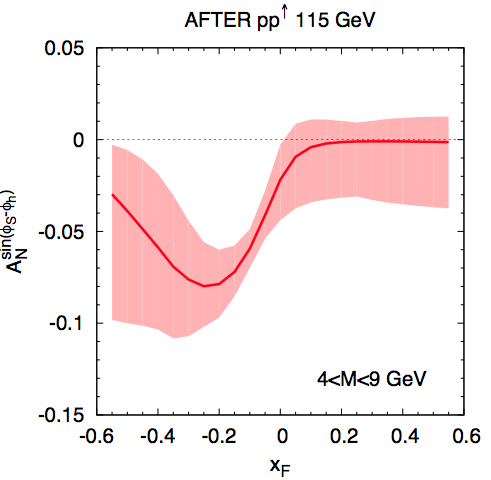} 
  \caption{Expected Sivers asymmetry in Drell-Yan measurement at AFTER (no scale evolution). Plot taken 
from~\cite{talk-Mauro}, courtesy of U. D'Alesio. \label{fig:SSA-DY}}
  \vspace{0.43cm}
 \end{minipage}
\end{figure}

\section{Quark distribution at large $x$}
\label{sec:light}

The importance of DY data from fixed-target experiments in global PDF fits is well recognised. 
By measuring with AFTER DY pair production in both $pp$ and 
$pd$ collisions in the backward region, the antiquark distributions, 
$\bar{u}(x)$ and $\bar{d}(x)$, in the nucleons can be accessed at rather low $x$ , complementing 
the forthcoming studies by E906~\cite{Reimer:2011zza}. 

In singly  polarised proton-proton collisions, it is also a sensitive probe of 
the quark Sivers effect and thus of the correlation between the nucleon spin and 
the transverse momentum $k_T$ of the quark in the nucleon. The study of such correlations 
are invaluable in the quest to understand the structure of the spin of the nucleons. It has been 
shown~\cite{Liu:2012vn} that AFTER is well positioned to measure the quark Sivers effect in DY pair production 
via such Single-transverse-Spin-Asymmetry (SSA) studies. 
The domain covered by AFTER in $pp^{\uparrow}$ collisions can basically  go up to $x_F \rightarrow -1$. 
In this region, the SSA are sensitive to partons with large momentum fractions in the polarised 
nucleon, $x^{\uparrow }$. Taking into account the uncertainties obtained in~\cite{Anselmino:2013rya}, 
one expects DY asymmetries in the backward region up to 10\% in the mass region $4 < M < 9$ GeV (see \cf{fig:SSA-DY}).

\section{Gluon distribution at large $x$}
\label{sec:gluon}

At high energy, heavy quarkonium production proceeds through $gg$ fusion, whereas 
isolated photons originate from $gq$ fusion\footnote{A recent survey of isolated photon data and their impact on the determination
of the gluon PDF can be found in~\cite{d'Enterria:2012yj}.}. Both can be used as experimental tools to constrain the gluon PDF 
and nPDF at large $x$, which are poorly known (see {\it e.g.} \cf{fig:eps09}). A {\it sine qua non} 
condition to reach the large $x$ region with these rare probes is a high luminosity, which is precisely 
what a fixed-target experiment such as AFTER can easily provide~\cite{Brodsky:2012vg,Lansberg:2012kf} in 
$pp$ and $pA$ collisions. This implies that AFTER should be designed as a multi-purpose detector, 
capable of studying quarkonium and isolated photon production.  
While the use of quarkonia as gluon probe (see \eg~\cite{Diakonov:2012vb}) relies on a better understanding of their production mechanisms~\cite{review}, 
these would be better constrained thanks to the large quarkonium yields and precise measurements of their correlations, 
along with the forthcoming LHC results.

In addition, it will be particularly interesting to investigate the gluon content of the neutron and to see 
whether there is any deviation from that of the proton. A pioneering measurement was done by E866~\cite{Zhu:2007mja} at 
Fermilab, using \upsi. 
A unique opportunity to get significant improvements can be offered by AFTER, on two different sides: the precision and 
also the extension to lower $x$ and $Q^2$ values by using the \jpsi.  

Another question concerns the gluon momentum tomography in the nucleon. 
Is there any Sivers effect for the gluon? Is there any correlation, pinned down by the Boer-Mulders 
effect, between the gluon $k_T$ and the gluon linear polarisation?

A first attempt to measure a SSA  arising from the gluon Sivers effect  was recently carried out by the PHENIX collaboration~\cite{Adare:2010bd} by looking at \jpsi\ production in $pp^{\uparrow}$ collisions. 
Improvements are needed. On top of larger luminosities (by orders of magnitude), AFTER could offer to extend the 
measurement in various experimental probes sensitive to gluons, such as other 
quarkonium species (\upsi, $\chi_c$, $\ldots$), $B$ and $D$ meson~\cite{Anselmino:2004nk} production, 
prompt photon, photon-jet~\cite{Bacchetta:2007sz} or double photon production~\cite{Qiu:2011ai}.
Concerning the Boer-Mulders effect, linearly polarised gluons inside unpolarised protons~\cite{Boer:2012bt} can be accessed
with the study fo low-$P_T$ scalar and pseudoscalar quarkonium\footnote{See~\cite{Pisano:2013cya} for a discussion 
of the Boer-Mulders  effect in heavy flavour production in $pp$ collisions.}.

AFTER could also greatly contribute to the knowledge of the gluon nPDF in \pA\ collisions, in good complementarity 
with LHeC~\cite{AbelleiraFernandez:2012cc} (focusing at low $x$) and EIC future facilities~\cite{Accardi:2012hwp} (at intermediate $x$). On top of the $x$ dependence, 
a more precise $A$ dependence is needed. It is easier in a fixed-target setup than in the collider mode 
thanks to the target versatility. Such results can provide a much better evaluation of the nuclear matter effects 
on quarkonium and heavy-flavour production, which is a key milestone towards any precision study of the deconfinement at 
RHIC energies.

\section{Heavy quark distribution at large $x$}
\label{sec:heavy}
 
The presence of an intrinsic charm component in the nucleon  is still the subject
of intense debates in the particle and hadron physics community.  At large $x$, 
whereas compelling indication of intrinsic strangeness has been recently claimed to be found~\cite{Chang:2011vx},
a state-of-the-art global fit~\cite{Pumplin:2007wg} has shown
that various realisations of intrinsic charm PDFs are allowed by the data. Similarly, the existence of 
intrinsic bottom is surely worth some investigations (see \eg~\cite{Brodsky:2012yy}.) 

To advance the debate, dedicated measurements are absolutely needed; not only one. Several probes are accessible 
at AFTER~\cite{Brodsky:2012vg,Lansberg:2012kf}, 
such as backward open charm and beauty production, double quarkonium production, quarkonium plus open heavy flavour production, as well as the production of prompt photon in association with a heavy quark~\cite{Stavreva:2009vi,Stavreva:2010mw}.
All these measurements could also be completemented by a new measurement of the charm structure function in DIS.

\section{Summary}

A novel testing ground for QCD can be provided by AFTER, a fixed-target facility based 
on the multi-TeV proton or heavy-ion beams at the LHC extracted by a bent crystal, interacting with a 
fixed proton, deuteron or nuclear target, possibly polarised. 
Such a fixed-target mode gives access to outstanding luminosities  
at c.m.s. energies comparable to RHIC energies. For the first time, AFTER provides a systematic access of the 
largely uncharted region at $x_F \rightarrow -1$ or large $x$ in the target, thus opening high accuracy studies 
of the quark and gluon content of the nucleon in this region.

{\bf Acknowledgements.}
This research [SLAC-PUB-15718] was supported in part by the French CNRS, grants PICS-06149 Torino-IPNO \& PEPS4AFTER2, 
the Department of Energy, contract DE-AC02-76SF00515 and the Sapore Gravis networking 
of the I3 Hadron Physics program of the EU 7th FP.

\end{document}